\documentclass[aps,prl,twocolumn,groupedaddress,showpacs]{revtex4-1}

\usepackage{graphicx}
\usepackage[geometry]{ifsym}
 \usepackage{epsfig}
 \usepackage{color}
 \usepackage{soul}
 \usepackage{amsmath}
\begin{document}

\title{High-spin polaron in lightly doped CuO$_2$ planes}

\author{Bayo Lau} \author{Mona Berciu} \author{George A. Sawatzky}
\affiliation{Department of Physics and Astronomy, University of
  British Columbia, Vancouver, BC, V6T 1Z1}

\date{\today}

\begin{abstract}
We derive and investigate numerically a minimal yet detailed spin
polaron model that describes lightly doped CuO$_2$ layers. The
low-energy physics of a hole is studied by total-spin-resolved exact
diagonalization on clusters of up to 32 CuO$_2$ unit cells, revealing
features missed by previous studies. In particular, spin-polaron
states with total spin 3/2 are the lowest eigenstates in some regions
of the Brillouin zone. In these regions, and also at other points, the
quasiparticle weight is identically zero indicating orthogonal states
to those represented in the one electron Green's function. This
highlights the importance of the proper treatment of spin fluctuations
in the many-body background.
\end{abstract}

\pacs{71.10.Fd,75.10.Jm,71.38.-k,74.72.-h}

\maketitle

{\em Introduction:} A full understanding of the physics of a CuO$_2$
layer doped with a few holes has still not been achieved,
despite continuous effort~\cite{dump,stripe}.  Recent high resolution angular resolved
photoemission (ARPES) studies \cite{damascelli,ronning,shen} on the
insulating charge-transfer gap parent compounds~\cite{zsa} reveal major puzzles:
do quasiparticles of one electron nature exist and if so what is their
energy and momentum? Why are the first visible electron removal states
so broad, of 300 meV at 300K and decreasing linearly with
temperature $T$, and what causes the very apparent $T$-dependent
change of line shapes? Is the momentum dependence of the lowest energy
structure related to the pseudo-gap formation at higher hole
concentrations? Recent neutron experiments performed in
the pseudo-gap phase reported magnetic response throughout
the Brillouin zone, not restricted to the region of the
much discussed magnetic resonance \cite{greven}. These and other issues including the
broken local 4-fold symmetry, which is taken for granted
in single-band models, seen in scanning tunneling probe (STM)\cite{davis}
and X-ray scattering~\cite{xray} remain
either open questions or are controversial.

It is widely believed that a complete description of a single hole
  in a spin-${ 1\over 2}$ 2D antiferromagnet (AFM) with {\em full} quantum fluctuations
  could provide the answers to these questions, as well as clues for
  understanding the origin of the non-Fermi-liquid behavior and the
  superconducting ground-state in the higher hole density region. Of
  course, consideration of exotic many-body scenarios \cite{varma}, or of
  coupling to lattice vibrations
  \cite{phonon} are exciting developments; however, a detailed modeling
  of the hole+AFM is a crucial first step to understand the
  significance of such additions. This problem is very difficult
  because of the complicated nature of the
  2D AFM background, whose quantum fluctuations
  in the presence of doped holes were never fully captured for a large CuO$_2$ lattice.  Recent
technical developments \cite{octapartite} allow us to present the
first such results for samples with up to $32$ Cu and $64$ O, in this
Letter. Our work also reveals important failings of the single-band models
used extensively in the literature, and which we briefly review
below~\cite{footnote0}.

Microscopic hole-AFM interactions have been studied in models with one
\cite{zrs,russianpolaron,tjrev,tjed}, two
\cite{zaanen2band,emery1,macridin2band}, three \cite{threeband}, or more \cite{aharony,beyond} bands. While exact
analytical solutions seem to be out of reach, numerical studies are
always carried out with compromises such as the use of small clusters and
variational approaches
\cite{numrev}.  Given these difficulties and the drive to find the
simplest model, the one-band models are unsurprisingly the
most studied \cite{tjrev}. While certain higher-energy aspects observed by XAS, EELS
and STM~\cite{details,stripe} cannot be described using one band, the
significance of omitting other bands in the low-energy scale cannot be quantified without a
comparison to unbiased solutions of more detailed models.

Cuprates exhibit charge-transfer band-gap behavior with
mobile holes located mainly on anion ligands and unpaired electrons on
cation $d$-orbitals \cite{zsa}.  One-band models use superexchange
\cite{superx} and Zhang-Rice singlets (ZRS) \cite{zrs} to reduce the
$(N-n)$-electron problem to one of $n$ holes in an AFM
background, often modeled as a N\'eel background with
spin-waves. To reach agreement with experiments, such models must
be tweaked at least by adding longer-range
hopping~\cite{tp,macridin2band}.  One trade-off for their
elegance is the use of momentum-independent effective parameters,
even though it is well known that the ZRS state has a strong
$\mathbf{k}$-dependent renormalization~\cite{zrs}. The  impact of such approximations
must be verified for all $\mathbf{k}$ with models
that distinguish anion and cation sites.

In this Letter we study a single hole in a  model that
includes the O $2p$ orbitals explicitly and {\em full quantum fluctuations} of the AFM
background. This results in spin-polaron solutions absent from other
models or approximations used to date. The energy dispersion of the
lowest energy electron removal state is similar
 to that of the ZRS,
which effectively locks the O hole in a singlet with {\em one of the two} adjacent
copper sites. Without such restrictions, our
resulting wavefunction features the hole forming a stable $S={1\over 2}$
three-spin polaron (3SP) with its {\em two neighbour copper} sites. In some
regions of the Brillouin zone, a  $S=1$ quantum fluctuation binds to
the $S={1\over 2}$ polaron yielding a low-energy $S={3\over 2}$ state
invisible at $T=0$
to ARPES. For all momenta, the $S={3\over 2}$ states are found to be within $\lesssim
J/2$ of the $S={1\over 2}$ band. These and other
results inconsistent with one-band models, and their
experimental implications, are discussed below.

We start with the three-band $p-d$ model which exhibits the basic
physics of a hole doped charge transfer gap and insulating spin 1/2
antiferromagnet \cite{threeband}:
\begin{eqnarray}
H_{3B}&=&T_{pd}+T_{pp} + \Delta_{pd}\sum n_{l+\epsilon,\sigma}\nonumber\\
&&+U_{pp}\sum
n_{l+\epsilon,\uparrow}n_{l+\epsilon,\downarrow} +U_{dd}\sum n_{l,\uparrow}n_{l,\downarrow}\label{1}
\end{eqnarray}
where $n_{l,\sigma}=d^\dag_{l,\sigma} d_{l,\sigma}$,
$n_{l+\epsilon,\sigma}=p^\dag_{l+\epsilon,\sigma}
p_{l+\epsilon,\sigma}$ count holes in  Cu
$3d_{x^2-y^2}$, respectively
O $2p_{x/y}$ orbitals (see Fig. 1) and  $U_{dd}>U_{pp}>\Delta_{pd}$
describe Hubbard and charge transfer interactions. Nearest neighbor (NN) Cu-O hopping
 $T_{pd}= t_{pd}\sum
[(p^\dag_{l+\epsilon,\sigma}-p^\dag_{l-\epsilon,\sigma}) d_{l,\sigma}
  +h.c.]$ is included, as is  hopping  $T_{pp} = t_{pp}\sum
s_\delta p^\dag_{l+\epsilon+\delta,\sigma}
p_{l+\epsilon,\sigma}-t_{pp}'\sum
(p^\dag_{l-\epsilon,\sigma}+p^\dag_{l+3\epsilon,\sigma})
p_{l+\epsilon,\sigma}$ between NN and certain NNN O sites. For NN
hopping by $\delta=(\delta_x,\delta_y)$, $s_\delta=\delta_x\delta_y/|\delta_x\delta_y|$.


In a half-filled, large-$U$ system with \emph{no hopping}, the
ground-state (GS) has a hole at each Cu site: $\prod
d^\dag_{l,\sigma_l}|0\rangle=\prod |\sigma_l\rangle$, with the usual
$2^N$ spin degeneracy. An electron removal adds a hole in an O
orbital, so the doped GS is $p_{l+\epsilon,\sigma}^\dag \prod
|\sigma_l\rangle$, with $2N\times2^{N+1}$ degeneracy. We study the
 behavior of such anion holes when the hopping is turned on, in the
 framework of superexchange. The
idea is reminiscent of studies such as
Refs. \cite{zaanen2band,emery1,aharony}; however, these also used further
approximations. A detailed comparison of our  Hamiltonian
versus those used in these references is provided in the Supplementary
Material~\cite{footnote0}.

\begin{figure}
\includegraphics[width=\columnwidth]{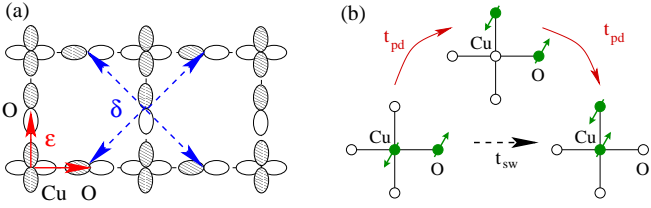}
\caption{\label{fig:cuo2} (a) Two adjacent unit cells of the CuO$_2$
  plane. The orbitals kept in the 3-band model of Eq. (\ref{1}) are
  shown, with white/shaded for positive/negative signs. The two
  $\epsilon$ vectors (solid arrow) and the
  four $\delta$ vectors (dashed arrow) are also shown. (b) Sketch of
a virtual process of $T_{\rm swap}$.}
\end{figure}

{\em Model:} Noting that all $T_{pd}$ processes increase energy by
either $U$ and/or $\Delta_{pd}$,
 we derive the effective model for the states $p_{l+\epsilon,\sigma}^\dag \prod
|\sigma_l\rangle$ to be~\cite{footnote0}:
\begin{equation}
H_{\rm eff}=T_{pp}+T_{\rm swap}+H_{J_{pd}}+H_{J_{dd}}\label{eq:heff}
\end{equation}
where the O-O hopping of the hole is supplemented by:
\begin{eqnarray}
T_{swap}&=&-t_{sw} \sum s_{\eta}p^\dag_{l+\epsilon+\eta,\sigma}p_{l+\epsilon,\sigma'}
|\sigma'_{l_{\epsilon,\eta}}\rangle\langle\sigma_{l_{\epsilon,\eta}}|\label{eq:tex}\\
H_{J_{pd}}&=&J_{pd} \sum\overline{S}_{l}\cdot\overline{S}_{l\pm\epsilon}\label{eq:jpd}\\
H_{J_{dd}}&=&J_{dd}\sum\overline{S}_{l\pm2\epsilon}\cdot\overline{S}_l\Pi_\sigma
(1-n_{l\pm\epsilon,\sigma})\label{eq:jdd}
\end{eqnarray}
Using $t_{pd}=1.3eV$,
$t_{pp}=0.65eV$, $t_{pp}'=0.58t_{pp}$, $\Delta_{pd}=3.6eV$, and
$U_{pp}=4eV$ \cite{tjrev}, we scale the parameters in units of
$J_{dd}$ to find their {\em dimensionless} values to be $t_{pp}=4.13$,
$t_{sw}=2.98$, and $J_{pd}=2.83$.

While we find that the 3-spin polaron (3SP)~\cite{emery1}
plays an important role, our
approach is different from previous work~\cite{zaanen2band}
by recognizing (i) $T_{pp}$'s role as a coherence facilitator rather than a
mere correction; (ii) its complementing process $T_{swap}$,
illustrated in Fig. 1(b),
(iii) suppression of superexchange along the bond inhabited by the hole,
see Eq. (\ref{eq:jdd})~\cite{footnote0}, and iv) total-spin ($S_T$) eigenstates are studied explicitly. We push the
computational limit to perform
$S_T$-resolved exact diagonalization (ED) of a topologically
superior \cite{betts} cluster of $N=32$ CuO$_2$ unit cells,
treating the AFM background exactly. ED provides the transparency,
flexibility, and neutrality to support new results. The price for a
systematic mapping of the excited states is the limited $\mathbf{k}$ resolution. $N=16$
results are provided to check for size dependence.

We find that all low-energy eigenstates have a total spin of either
$S_T={1\over 2}$ or ${3\over 2}$. The $z$-projections for each $S_T$ are
degenerate. The $S_T={1\over 2}$ subspace is due to the $s={1\over 2}$
hole mixing with various $S=0$ background states, including the AFM GS, or mixing with
the $S=1$ background states, including the ``single-magnon'' states. The $s={1\over 2}$
carrier can also mix with $S=1$ or 2 background states to yield the
$S_T={3\over 2}$ subspace;
we explicitly consider such states here for the first time.  The partition
of the
$S_T^z$ subspace into separate $S_T$ sectors was managed
by the optimizations of Ref. \cite{octapartite}. Unlike
there, \emph{no basis truncation} was employed here for rigorous
results. The ({$\mathbf{k}, S_T={3\over 2},S_T^z={1\over 2}$) sector
contains $\sim$0.44$\times10^9$ states.

\begin{figure}[t]
\includegraphics[width=0.8\columnwidth]{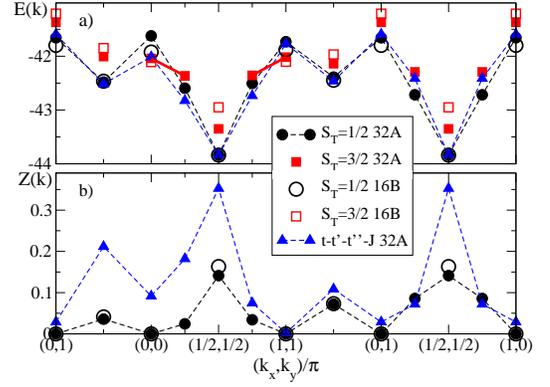}
\caption{\label{fig:fulle} a) Energy and b) quasiparticle weight
  (bottom) for the lowest eigenstates with $S_T={1\over 2}$ and
  ${3\over2}$ vs. momentum. Different sets are
  shifted so as to have the same GS energy.}
\end{figure}

\begin{figure}[t]
\includegraphics[width=0.70\columnwidth]{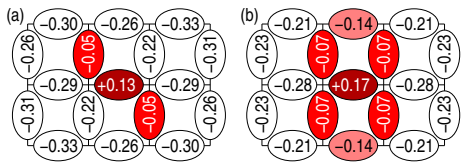}
\caption{\label{fig:w0_05_05} $\langle C_x(\delta,a)\rangle$ for the lowest energy
  state at (a) $(\frac{\pi}{2},\frac{\pi}{2})$ with
  $S_{T}={1\over 2}$, and (b) at ($\pi$,$\pi$) with
  $S_{T}={3\over 2}$. The darkly-shaded bullet denotes the oxygen
  position at $l+e_x$. Each bullet shows the correlation value between
  the two sandwiching Cu sites. The central 12 Cu sites are shown; the
correlations between the other 20 Cu spins converge fast towards the
AFM value of $\sim$-0.33. $\langle C_y(\delta,a)\rangle$ is the $\hat{P}_{x\leftrightarrow y}$ reflection.}
\end{figure}

{\em Results:} Fig.~\ref{fig:fulle}(a) shows the lowest eigenenergies.
The GS  has $\mathbf{k}=(\frac{\pi}{2},\frac{\pi}{2})$ and
$S_T={1\over 2}$, is consistent with the 3SP
but can also be thought of in terms of ZRS
~\cite{footnote0}.
Remarkably, we find similar
dispersion along (0,0)$\rightarrow$($\pi$,$\pi$) and
(0,$\pi$)$\rightarrow$($\pi$,0) without having to add longer range
hopping or fine-tune parameters as is needed in one band models. The
biggest surprise, though, are the low-lying $S_T={3\over 2}$ states
which go below the ${1\over 2}$ states near (0,0) and
($\pi$,$\pi$). Finite-size analysis, discussed in the Supplementary Material, reveals that
the $S_T={3\over 2}$ states are stable polarons at least in the
regions marked by thick solid lines in Fig. \ref{fig:fulle}(a). Thus,
\emph{ a $S_T={1\over 2}$ quasiparticle cannot describe the low-energy
  states throughout the BZ}.  To compare the lowest energy states on
both sides of the crossing, we note that the lowest $k_x=k_y$
eigenstates have odd parity upon a $\hat{P}_{x\leftrightarrow y}$
reflection (Fig.\ref{fig:cuo2}a) so they can be expressed as
$2^{-1/2}(1+\hat{P}_{x\leftrightarrow y})\sum
e^{ikl}p^\dag_{l+e_x,\sigma}|\sigma,l\rangle_x$.  The band-crossing
results in noticeable change in the expectation values of the
correlation function:
\begin{equation}
\hat{C}_{x}(\delta,a)=2\sum_{l,\sigma}\overline{S}_{l+\delta}
\cdot\overline{S}_{l+\delta+a}n_{l+e_x,\sigma}
\end{equation}
which measures the correlation between two neighboring Cu spins
at a distance $\delta$ from the hole. $\langle C \rangle$ ranges from
-3/4 for singlet, to $\sim$-0.33 for 2D AFM, to 1/4 for triplet.

Fig. \ref{fig:w0_05_05}a shows $\langle \hat{C}_{x}\rangle$ when
the hole is located at the darkly shaded bullet, in the GS:
$\mathbf{k}=({\pi\over 2},{\pi\over 2}$), $S_T={1\over 2}$
($\langle\hat{C}_{y}\rangle$ is a reflection with
$\hat{P}_{x\leftrightarrow y}$ for $k_x=k_y$). The hole affects the
AFM order in its vicinity. Because of the hole-spin exchange
$H_{J_{pd}}$ and the blocked superexchange between the two Cu spins
neighboring the hole, these ``central'' spins have triplet
correlations, of $\sim$0.13. Also, $\langle H_{J_{pd}}\rangle\sim$
-0.9$J_{pd}$, showing that locally this is consistent with the 3SP
solution~\cite{footnote0}. More interesting are the correlations with
the other 3 neighbors of each of these central Cu spins: with two of
them, there are robust AFM correlations of $\sim -0.22$, while with
the third the correlation nearly vanishes (lightly shaded
bullet). This is counterintuitive if one views the system as a
fluctuating Ne\'el background, where a spin-flip would change the
spin-spin correlation to all four neighbors. Although the two central
Cu spins have $2\over 3$ weight in triplet configuration which is
hardly bi-partite, long-range AFM order cannot be automatically
discounted \cite{lro}. Indeed, the correlations we find are consistent
with such order, except for the zigzag of 3 bonds shown by shaded
bullets. This strange shape is dictated by the hopping mechanism. For
a Bloch wave, O-O hole hopping in the upper-left/lower-right direction
yields a phase shift of $e^{i0}/e^{i(k_x-k_y)}$ and hence constructive
interference if $k_x=k_y$. In contrast, hopping in the
upper-right/lower-left direction yields a phase shift of
$e^{ik_x}/e^{-ik_y}$, and the interference is scaled down by
$\cos(k_{x/y})$. In the GS, having a mixture of
singlets and triplets upper-left/lower-right to the O hole lowers
energy with the least disturbance to AFM order. Thus, the two outside
zigzag bonds are triplet "disturbance tails" pointing orthogonal to
the momentum direction.  This is very different from the ZRS, which freezes
a Cu spin by intraplaquette coherence with its four O sites.

Fig. \ref{fig:w0_05_05}b shows the correlation values when the
$S_T={3\over 2}$ polaron becomes the lowest energy state at
($\pi$,$\pi$). The results look similar at (0,0). $\langle
H_{J_{pd}}\rangle$ remains $\sim-0.9J_{pd}$,
but there are now four more heavily disturbed bonds.
This further supports this being  a stable polaron
with an extra magnon bound locally close to the O hole.
We stress here that this $3\over 2$ polaron is formed by a
spin disturbance \emph{around} the 3SP. This is very different
from the $S=\frac{3}{2}$ excitation local to $H_{J_{pd}}$ with energy
$+\frac{J_{pd}}{2}$~\cite{aharony}.

Fig.~\ref{fig:fulle}b shows the quasiparticle weight $Z(\mathbf{k})$
for the first electron removal state. The major difference from other
models is that $Z(\mathbf{k})=0$ in three regions: a)
$Z(0,0)=Z(\pi$,$\pi$)=0 because here the lowest eigenstate has
$S_T={3\over 2}$ which due to spin-conservation is not in the
Krylov space of any $S_T={1\over 2}$ state~\cite{footnote0}, and b) $Z(0,\pi$)=0
even though this is a $S_T={1\over 2}$ state (see below).
The t-t'-t''-J model treatment
does not conserve $S_T$,
resulting in $Z(0,0)\sim0.1$ and a finite
$Z(\pi$,$\pi$)~\cite{tjed}.
Our $Z(k)$ is smaller everywhere
than that of the t-t'-t''-J model, suggesting less "free particle"
nature of the polaron.

The lowest energy state at $\mathbf{k}=(0,\pi)$ has $S_T={1\over 2}$, but
its $Z=0$ because
the state is not in the Krylov space of an electron
removal~\cite{footnote0}.
This seems to be due to symmetry, although we do not yet fully
understand this.
States with finite but small Z are at least 0.006J higher in
  energy. Their close existence above the $Z=0$ lowest
  state may be related to pseudogap phenomena in this
  region; however, this needs to be investigated in more detail.
Fig.~\ref{fig:w0_00_10} shows
the correlation for this $Z=0$ lowest state. Compared to the
GS (see Fig. \ref{fig:w0_05_05}a), there are
2 more disturbed bonds as required by the reflection parity about k. This larger
disturbance range is accompanied by more negative (AFM) correlation
values.

\begin{figure}[t]
\includegraphics[width=0.73\columnwidth]{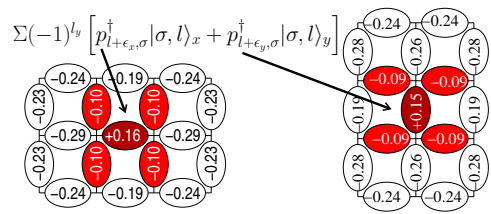}
\caption{\label{fig:w0_00_10} $\langle C_{x/y}(\delta,a)\rangle$
  for lowest state at k=(0,$\pi$) with $S_{T}$=1/2.}
\end{figure}

Although we are restricted to rather low momentum resolution, more can
be said about the $E_{\frac{3}{2}}-E_{\frac{1}{2}}=0$ band crossings
in the nodal direction by looking at the k points between which the
difference switch signs. The observation in Fig.~\ref{fig:fulle}a is
that, going away from the $(\frac{\pi}{2},\frac{\pi}{2})$ GS,
$E_{\frac{3}{2}}-E_{\frac{1}{2}}$ is larger towards $(0,0)$ than
towards $(\pi,\pi)$.  The ${1\over 2}\rightarrow{3\over 2}$ band
crossing would induce an \emph{abrupt change in Z(k) from non-zero to
  exactly zero, irrespective of the $Z(k)$ value on the finite
  side}. The larger $E_{\frac{3}{2}}-E_{\frac{1}{2}}$ towards
$k=(0,0)$ suggests that the non-zero region extends more towards
$k=(0,0)$ than towards $k=(\pi,\pi)$. Fig.~\ref{fig:fulle}a also shows
that the $3\over 2$ states get pushed further down as
$N\rightarrow\infty$ so \emph{the crossing is expected to be closer to
  the $(\frac{\pi}{2},\frac{\pi}{2})$ GS}. This is consistent with
ARPES which indeed observed an abrupt peak suppression in the nodal
direction as well as the peaks surviving longer towards
$k=(0,0)$~\cite{ronning}.

Even when the $S_T={3\over 2}$ states are not lowest in energy, they
hug the $S_T={1\over 2}$ band. This provides a $\lesssim J_{dd}$/2 energy scale for
spin excitations. At finite $T$, as magnons become
thermally activated, these ${3\over 2}$ states become ``visible'' to
ARPES. This suggests a $T$-dependent broadening mechanism of $\lesssim J_{dd}$/2 scale,
which, coincidentally, is the same energy scale recently linked to
phonons, as another possible source for this broadening~\cite{shen,phonon}.

Recent neutron experiments on samples at higher doping reveal $\sim 50meV$ magnetic
response centered at q=0, away the AFM resonance momentum~\cite{greven}. The bottom of
the single-particle band structure in Fig 2a) indeed has a q=0 1/2-to-3/2 excitation of this
energy scale. Our results so far are restricted to a single hole; nevertheless, it
has been pointed out that the q=0 magnetic excitation can be explained by
involving spins on oxygen sites~\cite{fauque}, which certainly are present in our results.

In addition to the low-energy 3/2 polaron
band, there are internal energy scales of the local 3SP since
$H_{J_{PD}}$ also has a $S={1\over 2}$ doublet and a $S={3\over 2}$ quartet
   separated in energy by $J_{pd}$ and 3$J_{pd}$/2 from the lowest energy state. Magnetic excitations of
  these energy scales have been observed via inelastic
  resonant X ray scattering even for doped samples without long-range AFM order~\cite{Keimer}.

{\em Summary:} We solved a detailed model which includes
the O sites and takes full account of the AFM  quantum fluctuations,
for large $N=32$ clusters.  The phases of the $p$ and
$d$ orbitals lead to phase coherence via
$T_{pp}+T_{swap}$~\cite{footnote0}.  This  is re-enforced by
$H_{J_{pd}}$ and the blocking of the AFM superexchange, making
corrections such as $T_{Kondo}$ negligible.
While the dispersion
  is similar to that measured by ARPES without any fine-tuning, the
  lifting of Cu-O singlet restriction present in ZRS-based models
  leads to wavefunctions of different nature, i.e. the 3SP where the
  O hole correlates with both its neighbour Cu sites. This model also
  provides low-energy channels for $S=1$ excitations.  $Z(k)$ was
found to be identically zero in certain regions of the BZ for two
reasons: 1) the spin-${3\over2}$ of the lowest energy state close to
$(0,0)$ and $(\pi,\pi)$; and 2) around the antinodal region because of
the lowest energy state there being exactly orthogonal to the single
electron removal state. The detailed nature of the state in the
anti-nodal region is still being investigated.

{\em Acknowledgement:} We thank G. Khaliullin for discussions, B. Keimer
for providing x-ray data, I. Elfimov and Westgrid for tech support, and CFI, CIfAR, CRC, NSERC
and Sloan Foundation for funding.


\end{document}